# XMM-Newton Observatory

D H Lumb[a], N Schartel[b] and F A Jansen[a]

Directorate of Science and Robotic Exploration, European Space Agency

[a]ESTEC, Postbus 299, 2200AG Noordwijk, Netherlands, +3171565446
david.lumb@esa.int; fred.jansen@esa.int

[b]Villafranca del Castillo, Apartado 78, E-28691 Villanueva de la Cañada, Madrid, Spain  +34 918131184 Norbert.Schartel@sciops.esa.int

Abstract:

X-ray Multi-Mirror Mission (XMM-Newton) has been one of the most successful astronomy missions launched by the European Space Agency. It exploits innovative use of replication technology for the X-ray reflecting telescopes that has resulted in an unprecedented combination of effective area and resolution. Three telescopes are equipped with imaging cameras and spectrometers that operate simultaneously, together with a coaligned optical telescope. The key features of the payload are described, and the in-orbit performance and scientific achievements are summarised.
Subject terms or keywords: XMM-Newton, X-ray astronomy, space telescopes

## 1 - Introduction

An X-ray astronomy mission for European Space Agency Horizon 2000 programme was studied from the early 1980's, culminating in a mission presentation at an ESA workshop held in Lyngby, Denmark in June 1985. In the papers presented at this conference [1] the mission design contained 12 low-energy and 7 high-energy telescopes with a collecting area of 13000 $cm^2$ and 10000 $cm^2$ at 2 and 6 keV respectively. The scientific goal was to maximise collecting area for spectroscopy, complementing the imaging science of the NASA AXAF programme. When the report of the telescope working group [2] was delivered in 1987, the consideration of practical constraints had reduced the number of telescopes to a more modest total of 7. The mission was approved into implementation phase in 1994, and an improved observing efficiency achieved with a highly eccentric orbit allowed the number of telescopes to be reduced. The development of suitable mirrors involved parallel studies of solid nickel, nickel sandwich and carbon fibre technologies up to (as late as) early-1995.  Soon afterwards the nickel electroforming replication technology was adopted, following the delivery of two successful mirror module demonstration models. The XMM flight model mirror modules were delivered in December 1998. XMM was launched on December 10, 1999, via the first commercial Ariane5 launch from Kourou, French-Guyana. It is the largest scientific European spacecraft to date; built and launched within the budget and schedule defined at approval. The overall mission cost was 689 MEuro (1999 economic conditions). In orbit it was renamed XMM-Newton.

A key aspect of the design was the simultaneous operation of 6 coaligned instruments, the three EPIC (European Photon Imaging Camera) imaging X-ray cameras, the two

RGS (Reflection Grating Spectrometer) grating X-ray spectrometers and the OM (Optical Monitor). We describe briefly the spacecraft configuration, followed by a more detailed description of the telescope design. Each of the instruments are presented, followed by a review of the in-orbit performance and the key science results achieved.

## **2** - Spacecraft

The configuration of the spacecraft is shown in an exploded view in Figure 1. It comprises 4 main sections:

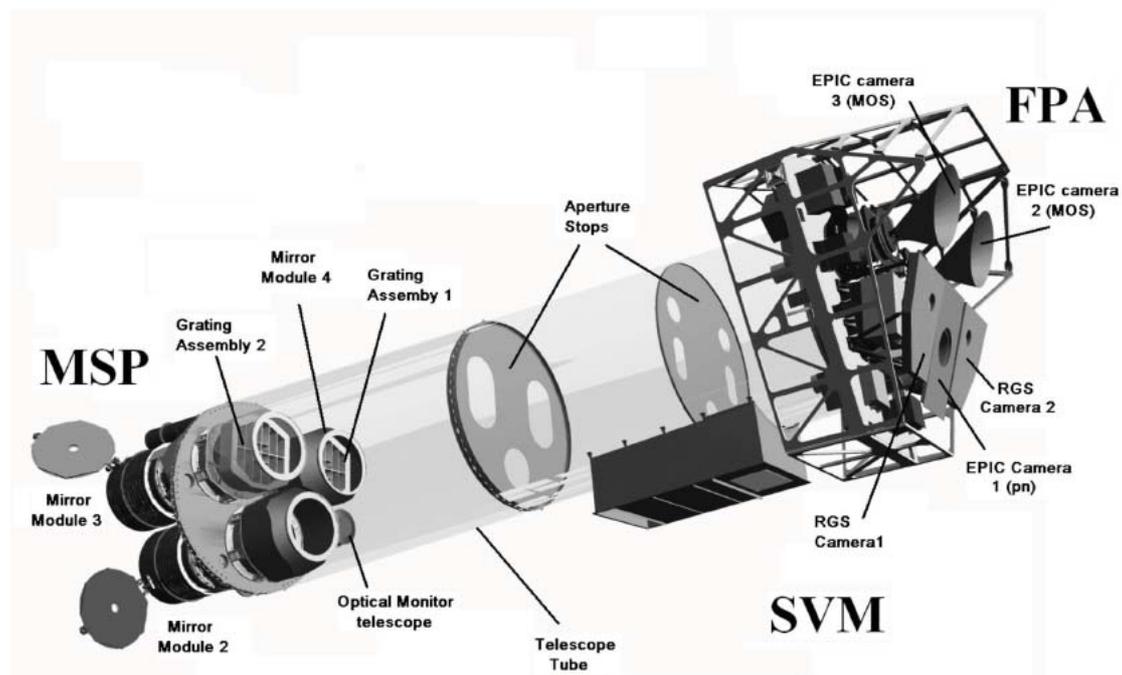

Figure 1 View of the XMM-Newton spacecraft subsystems, with external shrouds and structure removed for clarity.

The Focal Plane Assembly (FPA), consisting of a platform carrying the focal-plane instruments: two RGS readout cameras, an EPIC PN[1] and two EPIC MOS[2] imaging detectors, and electronics boxes. The EPIC and RGS instruments are fitted with radiators, to passively cool the CCDs (Charge Coupled Device)
The Mirror Support Platform (MSP), carrying the three mirror assemblies, the OM and the two star-trackers.
The long carbon fibre Telescope Tube (not shown in Figure 1), maintaining the FPA and the MSP separation. The upper half includes reversible venting and out-gassing doors and baffles.
The Service Module (SVM), carrying the spacecraft subsystems, the two solar-array wings, the Telescope Sun Shield (TSS) and the two S-band antennae.
The satellite was built under contract to ESA by a consortium of 35 European companies with Astrium (formerly Dornier) as prime contractor. The satellite mass is 3,800 kg and is ~10 metres long. XMM-Newton was placed in an eccentric 48 hour

---
[1] PN – Refers to the implanted junction semiconductor technology used in the detector production
[2] MOS – Refers to Metal-Oxide-Semiconductor technology used in the detectors production

elliptical orbit; at its apogee the altitude is ~114,000 km, and the initial perigee was only ~7,000 km. For reasons largely related to the mission design lifetime XMM-Newton has no onboard data storage capacity, so all data are immediately downloaded to the ground in real time, as facilitated by the ground station visibility. Contact for continuous real-time interaction with the spacecraft over almost the entire orbit is provided by the ESA ground stations at Perth and Kourou (with Villafranca and Santiago as back-up). Mission operations are managed through the ESA Spacecraft Operations Centre in Darmstadt, Germany. The observations are managed and archived at the European Space Astronomy Centre (formerly known as VILSPA) at Villafranca, Spain. The science data are processed at the XMM-Newton Survey Science Centre at the University of Leicester, England.

### 3 - Telescope

The three telescopes each consist of the following elements, as shown in Figure 2:

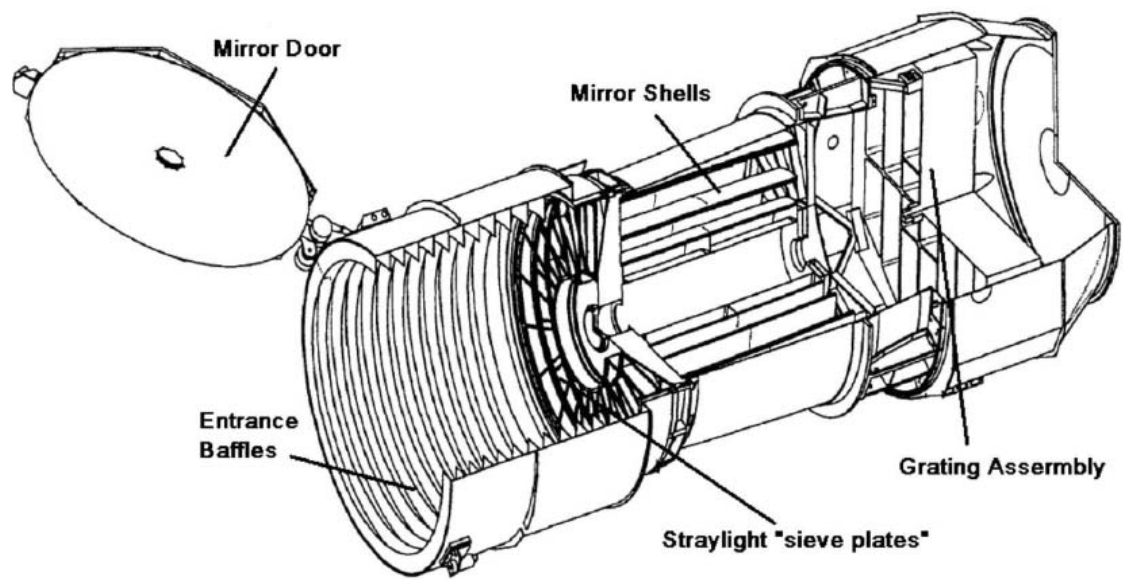

Figure 2: XMM-Newton telescope assembly

i) Mirror Assembly Door, which closed and protected the X-ray optics and the telescope interior against contamination until operations began.
ii) Entrance Baffle, which provides the visible stray light suppression
iii) X-ray Baffle (XRB) which blocks X-rays from outside the nominal field of view, which would otherwise reflect once on the hyperboloid section of the mirrors and would therefore cause stray light.
iv) Mirror Module (MM) the X-ray optics themselves.
v) Electron Deflector (producing a toroidal magnetic field for diverting "soft" electrons), right behind the mirrors in the shadow of the MM spider
vi) Reflection Grating Assembly (RGA), with a mass of 60 kg, only present on the backside of two out of three MMs. It deflects roughly half of the X-ray light to a strip of CCD detectors (RGS), offset from the focal plane
vii) Exit Baffle, providing a thermal environment for the gratings

A Mirror Module is a Wolter 1 type grazing incidence telescope with a focal length of 7.5 metres and with a resolution of ~15 arcsec (on-axis Half Energy Width)[3]. Each consists of 58 nested mirror shells bonded at one end on a spider. Design details are summarised in Table 1

| **Focal length** | 7500 mm |
|---|---|
| **Resolution (0.1-12 keV)** | |
| Half Energy Width | 15 arcsec |
| FWHM | <8 arcsec |
| **Effective area (1keV)** | 1500 cm$^2$ |
| **Mirror diameter and thickness** | |
| Outermost | 700 mm / 1.07 mm |
| Innermost | 306 mm / 0.47 mm |
| **Mirror length** | 600 mm |
| **Packing distance** | 1- 5 mm |
| **Number of mirrors** | 58 |
| **Reflective surface** | Gold (250 nm layer) |
| **Mirror Module mass** | 420 kg |

.

Table 1 Summary design parameters of an XMM-Newton Mirror Module
[3] Off-axis performance is described in the XMM Users Handbook section 3 avaailable
http://xmm.esac.esa.int/external/xmm_user_support/documentation/uhb_2.9

The X-ray mirrors are thin monolithic gold-coated nickel shells. The mirror shell manufacturing [3] is based on a replication process, which transfers a gold layer deposited on the highly polished master mandrel to the electrolytic nickel shell, which is electroformed on the gold layer. The mandrels are made out of initially double conical aluminium blocks coated with Kanigen nickel and then lapped to the exact shape and finally super-polished to a surface roughness better than 4 A (0.4 nm). The process is represented graphically in Figure 3

In order to allow rapid testing of the individual mirror shells and integrated mirror modules a special vertical test facility (using a UV beam) was developed at the Centre Spatial de Liège in Belgium [4]. The X-ray testing of the integrated XMM-Newton mirror modules was performed at the Panter facility of the Max-Planck Institut für Extraterrestrische Physik near Munich, Germany [5].

The technological development and the management of the X-ray mirror programme have been detailed in references [6.7]

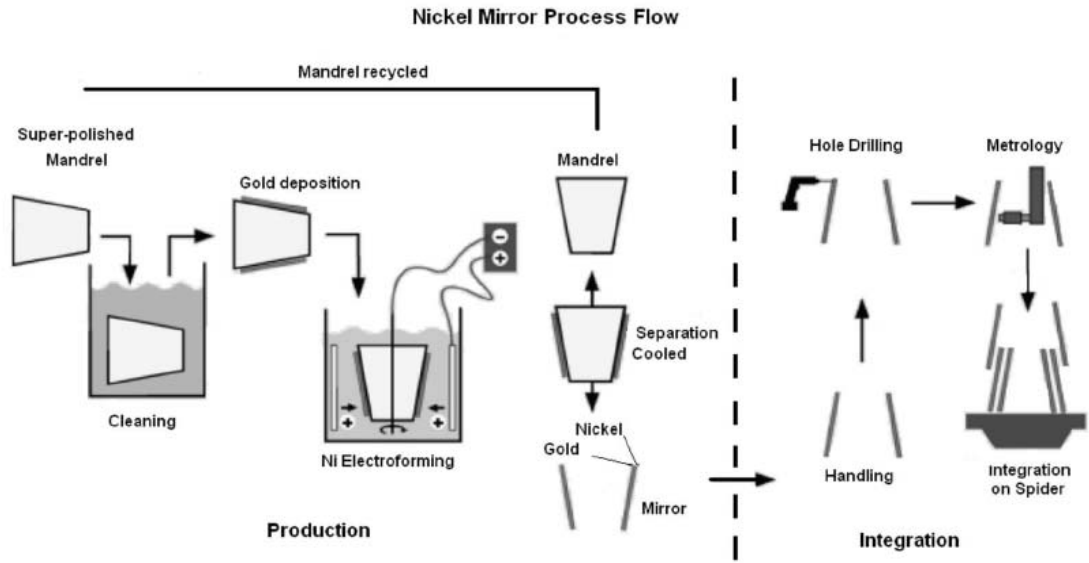

Figure 3 Mirror production process

## 4 - Instruments
### 4.1 EPIC

EPIC [8,9] comprises a set of three X-ray CCD cameras. Two of the cameras contain MOS CCD arrays (referred to as the MOS cameras). They are installed behind the X-ray telescopes that are equipped with the gratings of the RGS. The gratings divert about half of the telescope incident flux towards the RGS detectors such that about 44% of the original incoming flux reaches the MOS cameras. The EPIC instrument at the focus of the third X-ray telescope with an unobstructed beam; uses pn CCDs and is referred to as the pn camera. (Figure 4)

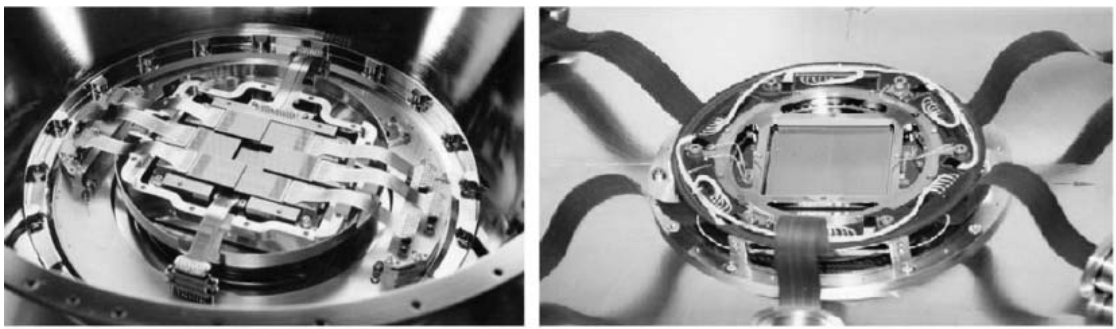

Figure 4 Flight assemblies of the EPIC cameras (Left - MOS camera Right – pn camera)

The EPIC cameras perform extremely sensitive imaging observations over the telescopes field of view (FOV) of 30 arcmin and in the energy range from 0.15 to 15 keV with moderate spectral ($E/\Delta E \sim 20\text{-}50$). All EPIC CCDs operate in photon counting mode with a fixed, mode dependent frame read-out frequency. '

*EPIC MOS CCDs*

The MOS CCDs, EEV type 22, have 600 x 600 pixels, each 40 microns square; they are frame-transfer devices and front illuminated [10]. One pixel covers 1.1" .The third phase electrode is open so that 40% of the area is only covered by 400 Å of silicon oxide. This improves the low energy response, giving appreciable sensitivity down to 150 eV. There being no buried channel makes this part of the pixel immune to channel damage such as that caused by soft protons interacting just below the surface of the pixel. This, combined with the rather thicker oxide and poly-silicon that covers the buried channel, contributes to the relative radiation hardness of the devices to soft protons.

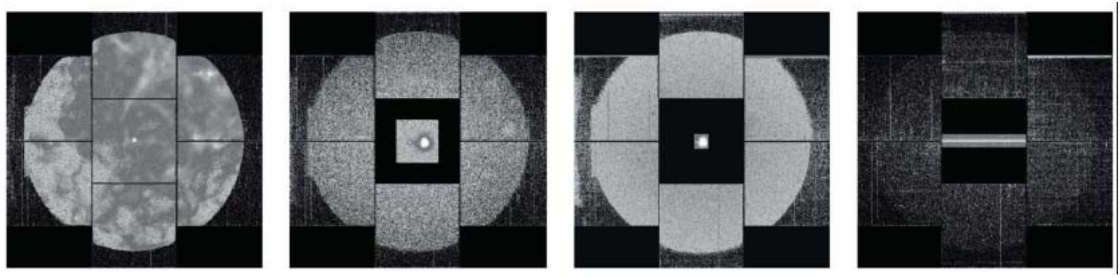

Figure 5 Representative images of the EPIC MOS camera acquired in the different operating modes (L-R Full Frame Image, Large Window, Small Window, Timing)

The full frame integration time is 2.6 seconds, governed by the time taken to read out the 600 x 600 pixels with sufficiently low noise (<5 electrons). The CCDs are read out in sequence.. The central CCD can be commanded independently of the others, which are commanded in pairs; the modes can be applied to the central CCD as described in Table 2 and Fig 5.

| CENTRAL CCD MODE | IMAGING AREA Pixels (arc min) | TIME RESOLUTION (s) |
|---|---|---|
| FULL | 600 x 600 (11 x 11) | 2.6 |
| LARGE WINDOW | 300 x 300 (5.5 x 5.5) | 0.6 |
| SMALLWINDOW | 100 x 100 (1.8*1.8) | 0.3 |
| TIMING | None (1.8 arc min projection) | ~$10^{-3}$ |

Table 2 Summary of EPIC MOS CCDs readout modes

The CCDs are cooled passively and maintained at their operating temperature using a three-stage radiator system combined with heaters. The minimum design temperature for the cooling system is –130 ° C. There is provision for de-icing the radiators and CCDs by raising the temperature to near 0°C and for annealing the CCDs by raising their temperature to +120 ° C.. Neither of these operations has so far needed to be performed. The nominal operating temperature of the CCDs at launch was –100 °C, and this is maintained to within ± 0.5 degrees by the control electronics.

### EPIC PN CCDs

The essential part of the EPIC PN detector is the 4 inch silicon wafer divided in an array of 12 monolithically implanted pn-CCDs, developed and manufactured in a dedicated semiconductor laboratory [11,12] . The array has a total effective size of 6 cm x 6 cm. The CCDs are arranged in four quadrants of three CCDs each. For redundancy reasons each quadrant can be considered as a separate unit: it has its own power supplies, back contact, preamplifiers and event analyzer electronics and can be operated independently from the others. The single CCD has a dimension of 3 cm x 0.98 cm with pixels of 150 μm x 150 μm arranged in 200 rows and 64 channels. The pixel size corresponds to an angular resolution element of 4.1 arcsec. The 64 read out anodes of each CCD are connected via on chip JFETs (Junction Field Effect Transistor) to a 64 channel charge sensitive amplifier (CAMEX64B). The 12 x 64 output channels of the CCDs as well as the supply and bias voltages are connected via bond wires to a multilayer printed circuit board, which carries the charge sensitive amplifiers as well as filter circuits for supply and bias voltages.

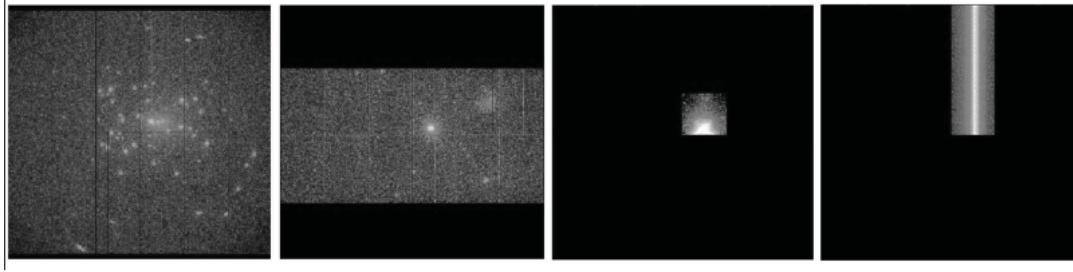

Figure 6 Readout modes of EPIC PN camera (L-R Full window, Large window, Small Window, Timing/Burst)

| CCD MODE | IMAGING AREA Pixels (arc min) | TIME RESOLUTION (ms) |
|---|---|---|
| FULL | 384 x 376 (26 x 27) | 73 |
| LARGE WINDOW | 384 x 198 (26 x 13.6) | 48 |
| SMALLWINDOW | 64 x 64 (4.4 x 4.4) | 5.7 |
| TIMING[a] | None (4.4 arc min projection) | 0.03 |

Table 3 Summary of EPIC PN CCDs readout modes. [a]An additional special version of TIMING mode provides even higher time resolution but with low duty cycle

Several read out modes can be selected for the pn-camera (Figure 6) to adjust the performance to the observation requirements (source flux). The full frame and window modes are the imaging modes. Timing and burst mode achieve better time resolutions at the expense of the coordinate in shift direction. In full frame mode, all

12 CCDs are read out sequentially within 70.25 ms, which corresponds to the time resolution in this mode. A CCD remains sensitive also during its read out cycle. This causes the so called 'out-of-time' events. These are events arriving during read out phase, when the charge content of the CCD is shifted to the read out anodes. The out-of-time events are distributed along the shift direction and get therefore a wrong coordinate in this direction. In full frame mode their fraction of 6.6% is given by the ratio between the read out time of one CCD and the read out time of all CCDs including 'warm up' times for preamplifiers.

### *EPIC Filter Wheel*

Because CCDs are light sensitive, two kinds of light filter are provided. Their design is a compromise between the need to prevent optical and UV photons from reaching the CCD plane, and the need to absorb as few X-ray photons as possible, especially at the lowest X-ray energy. The thin and medium filters comprise an unsupported polyimide film, 160nm thick, on with a single layer of aluminium; 40nm on the thin filter and 800nm on the medium. The transmission of visible light is approximately $10^{-2}$ and $10^{-4}$ respectively. The thick filter is constructed on an unsupported polypropylene film 330nm thick, with 160nm of aluminium and 40nm of tin to block the UV that would otherwise be transmitted by the polypropylene. Because of the large diameter (~65mm) and delicacy of these filters, it was felt unwise to launch them other than under vacuum, thus minimising the acoustic load, transferred to the filter from the rocket. Late in the campaign, a leak developed in one of the cameras that prevented the vacuum from being maintained in the long period between last access for pumping and the launch. This camera was filled with helium at atmospheric pressure with a blow-off valve to vent the camera during ascent. This system worked effectively to reduce acoustic load (cf. possible load with air) and the filters suffered no damage during launch. During radiation belt passage, and at all other times deemed unsafe for operation, the filter wheels are kept in the closed position.. There are 'Closed-Cal' observations used to monitor the long-term degradation of the CCD performance, by closing the filter wheel with a calibration source illuminating the CCDs. Once the radiation level is deemed to be safe, the filter wheel is moved to the appropriate filter position and the observation begins.

### 4.2 RGS Instrument

4.2.1. Optical design

The RGS design is illustrated in Figure 7. It incorporates an array of reflection gratings (RGA, aka grating stack) placed in the converging beam of an XMM-Newton telescope mirror module, this is done for two out of the three mirror modules on XMM-Newton.. The grating stack consists of 182 precisely aligned reflection gratings which in total intercept about half the light emanating from the telescope. The undeflected light passes through and is collected by the EPIC instrument in the telescope focal plane. The individual gratings are located on a toroidal Rowland surface, formed by rotating the Rowland circle about an axis passing through the telescope and spectroscopic foci, as illustrated in the left panel of Fig. 7. The gratings are slightly trapezoidal, since their edges lie along rays converging on the telescope focus.

While the field of view in the cross dispersion direction is determined by the width of the CCDs, the spatial resolution in this direction is largely determined by the imaging properties of the mirror. The situation in the dispersion direction is more complex as the source extent affects the wavelength resolution.

Nine large format back-illuminated CCDs are located on the Rowland circle to detect the dispersed spectra in single photon counting mode. The position of the X-ray on the detector is related to the X-ray wavelength through the dispersion equation. First, second and higher order spectra are overlapping on the detectors, but are easily separated by using the inherent energy resolution of the CCDs.

4.2.2. Reflection grating array

A Reflection Grating Array (RGA) contains 182 identical gratings. The gratings are mounted at grazing incidence in the in-plane, or classical configuration, in which the incident and diffracted X-rays lie in a plane that is perpendicular to the grating grooves. Because the beam is converging, the gratings are oriented so that the angle of the incident X-ray at the centre of the grating, α, is the same for all gratings in the array. In addition, the gratings all lie on the Rowland circle, which also contains the telescope focus and the spectroscopic focus for the blaze wavelength. In this configuration, aberrations, which would otherwise be introduced by the arraying, are eliminated [13]. The telescope aperture is filled by rotating the Rowland circle about an axis passing through the telescope and spectroscopic blaze foci. In all, each RGA contains six rows of gratings.

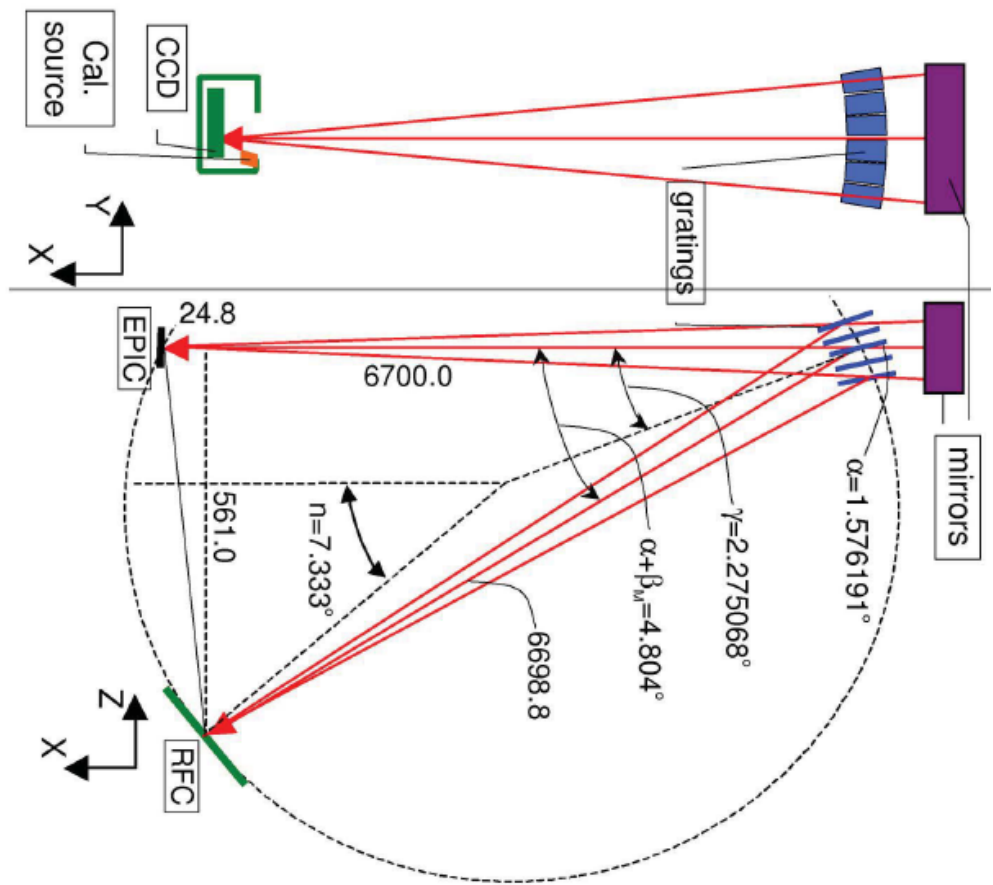

Figure. 7. Optical design of the RGS (not to scale). X-rays, indicated by red arrows, enter from the top. Numerical values for a few key dimensions and angles are indicated (linear dimensions in mm, angles in degrees).

Each grating measures about 10 by 20 cm (see Fig. 8). These large gratings need to be very flat in the long (i.e. dispersion) direction, since any non-flatness translates directly into a degradation of the resolution. At the same time the grating substrates need to be very thin in order to minimize the obstruction of the direct beam by the gratings. The grating substrates consist of 1mm SiC face sheets with five stiffening ribs at the back, running in the direction of the X-ray beam (see Figure 8). The face sheets are fabricated to 1 λ (634.8 nm) and 10 λ flatness in the long and the short direction, respectively. The gratings are replicated from a mechanically ruled master and are covered with a 200nm gold coating. The groove density varies slightly (± 10 %) over the length of the gratings, to correct for aberrations associated with the converging beam [14,13]. The groove density is ~646 grooves/mm at the centre.

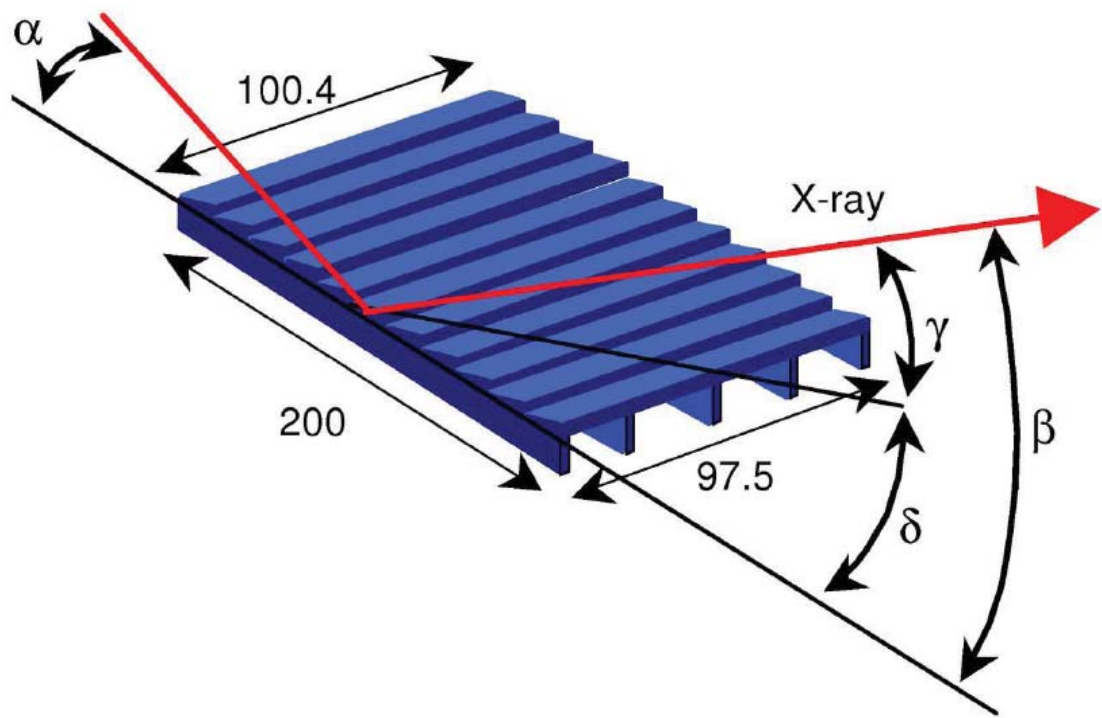

Figure. 8. Schematic drawing of a grating, including some of the key dimensions and angles (α=1.5762 , β=2.9739 (for blaze wavelength of 1.5 nm, γ=2.2751 , δ=0.6989, angles all in degrees).

The grating array support structure is made out of vacuum hot-pressed beryllium, which was selected for its low specific mass and good stability over the operational temperature range (10-30 ºC). To obtain the desired resolution, it is essential that all gratings are properly aligned (with 1 μm tolerance on the position of any grating corner). For a more detailed description of the RGA's see [15] .

4.2.3. RGS Focal plane camera

The dispersed spectrum is integrated on nine large format back illuminated CCDs. These nine CCDs are mounted in a row, following the curvature of the Rowland circle (see top two panels of Fig. 9). To reduce the dark current and improve general performance the CCD's are cooled to -80 ºC. An increase in CCD charge transfer inefficiency (CTI) due to radiation damage can be reduced by lowering the temperature to -120 ºC. Cooling is accomplished by a two-stage radiator, facing deep space. The CCD bench is housed internally to three nested thermal shells around the CCD bench [16] where the first shield also contains four internal calibration sources. The nine CCD chips are back-illuminated EEV devices with an image and storage section of 384 by 1024 pixels each and a pixel size of 27 μm by 27 μm and the length of the CCD assembly (253 mm) covers the first order dispersion of the 0.6 – 3.8 nm wavelength range. The instrument can be operated in a spectroscopy, high time resolution and diagnostic mode.

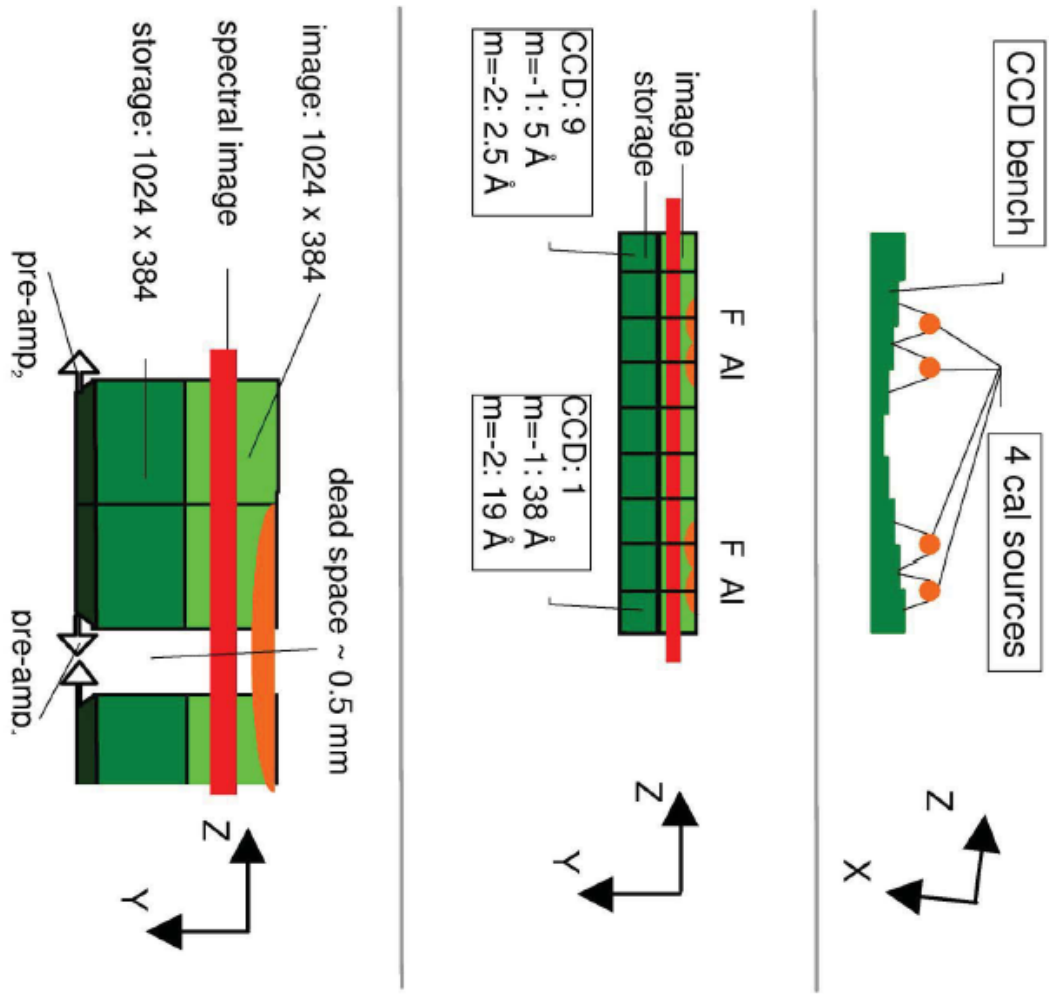

Figure. 9. CCD bench (top two panels) as well an enlarged view of two adjacent CCDs (bottom panel). The dead space indicated is present between each pair of CCD's. The figure is not to scale.

### 4.3 Optical Monitor

The Optical/UV Monitor Telescope (XMM-OM [17]) is mounted on the mirror support platform alongside the X-ray mirror modules. It provides coverage between 170 nm and 650 nm of the central 17 arc minute square region of the X-ray field of view, permitting routine multi-wavelength observations of XMM targets simultaneously in the X-ray and ultraviolet/optical bands A schematic of the XMM-OM is shown in Figure 10. It consists of two modules which are physically separated on the spacecraft; the telescope module and dual redundant digital electronics modules. The Telescope Module contains the telescope optics and detectors, the detector processing electronics and power supply. The Digital Electronics Module houses the Instrument Control Unit, which handles communications with the spacecraft and commanding of the instrument, and the Data Processing Unit, which pre-processes the data from the instrument before it is telemetered to the ground.

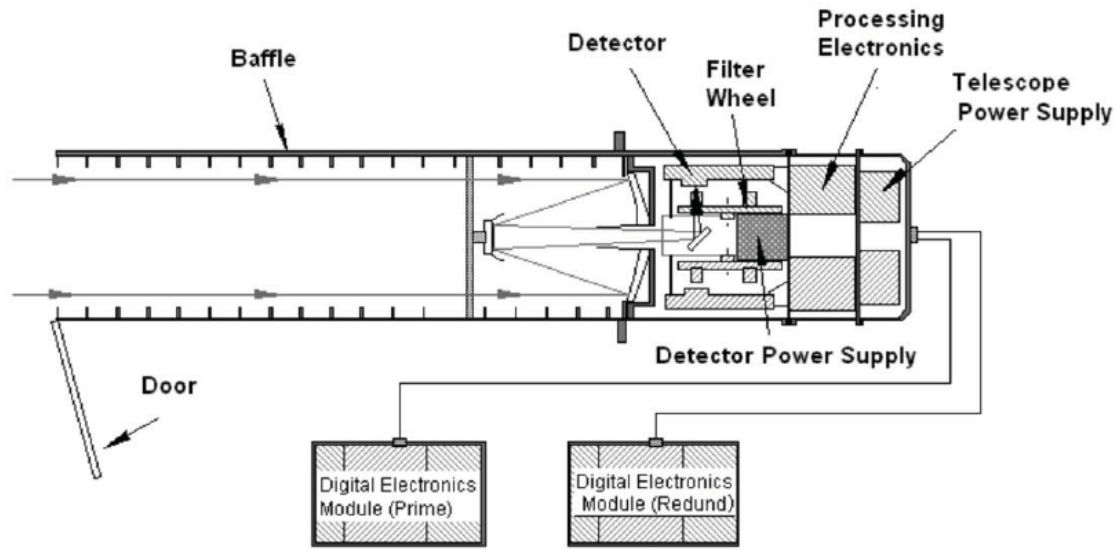

Figure 10 System block diagram of Optical Monitor

The telescope is of a Ritchey-Chrétien design and has a clear aperture of 30 cm. The f/ratio of the primary mirror is f/2.0, which is modified by the secondary to f/12.7, i.e. a focal length of ca. 3.8 m. The light beam is intercepted by a 45° flat mirror located behind the primary mirror which can be rotated to direct the beam to one of two redundant filter-wheel/detector assemblies. The format of the detector is 2048 x 2048 pixels with each pixel 9.5μm square. The field of view is 24 arcmin on the diagonal. In order to flatten the intrinsically curved focal plane, the detector window has been made concave (thinner at the centre), and the filters are weakly figured.

The detector is a microchannel plate (MCP)-intensified CCD (MIC). Incoming photons are converted into photoelectrons in an S20 photocathode deposited on the inside of the detector window. The photoelectrons are proximity focused onto a stack of three microchannel plates, which amplifies the signal by a factor of a million, through a bias of ~1.8kV. The resulting electrons are converted back into photons by a P46 phosphor screen. Light from the phosphor screen is passed through a fibre taper which reducing the image scale to compensate for the difference in physical size between the microchannel plate stack and the fast-scan CCD used to detect the photons.

In each detector there is a filter wheel. The filter wheel has 11 apertures, one of which is blanked off to serve as a shutter, preventing light from reaching the detector. Another seven filter locations house lenticular filters, six of which constitute a set of broad band filters for colour discrimination in the UV (UVW1, UVM2, UVW2) and optical (U, B, V) between 180 nm and 580 nm. The seventh is a "white light" filter which transmits light over the full range of the detector to give maximum sensitivity to point sources. (See Figure 11) The remaining filter positions contain two grisms, one optimized for the UV and the other for the optical range, and a x4 field expander

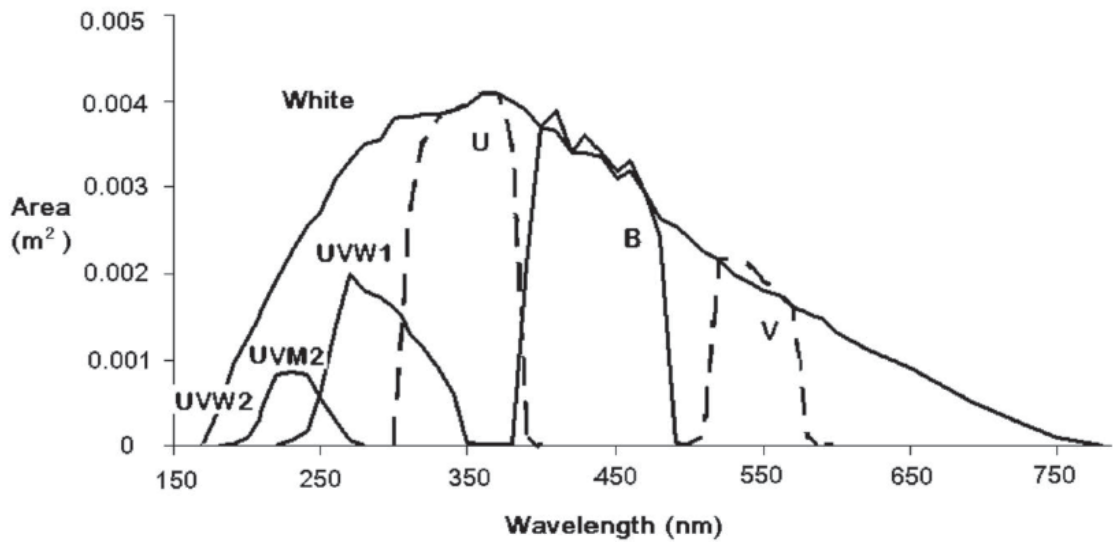

Figure 11 OM Filter transmissions

The signal from the CCD detector is processed by fast electronics near the detector head to extract information which is then transmitted to the data processing unit (DPU) in the digital electronics module (DEM). Each DEM contains an Instrumental Control Unit (ICU) and a Digital Processing Unit (DPU). The ICU commands the XMM-OM and handless communications between the XMM-OM and the spacecraft. The DPU is an image processing computer that digests the raw data from the instrument and applies a non-destructive compression algorithm ("tiered block word length" type) before the data are telemetered to the ground via the ICU. The DPU supports two main science data collection modes, which can be used simultaneously. The DPU autonomously selects up to 10 guide stars from the full OM image and monitors their position in detector coordinates at intervals that are typically set in the range 10-20 seconds, referred to as a tracking frame. These data provide a record of the drift of the spacecraft during the observation accurate to ~ 0.1 arc second.

The OM telescope module consists of a stray light baffle and a primary and secondary mirror assembly. . The separation of the primary and secondary mirrors is critical to achieving the image quality of the telescope. The separation is maintained to a level of 2 microns by invar support rods that connect the secondary spider to the primary mirror mount, and by maintaining an isothermal condition through distributing the detector electronics heat with heat pipes.

### 4.4 Instrument engineering challenges

On the instrument side, there were a number of main areas where challenging problems needed to be mastered.  In the areas related to instrument/detector development these most notably included:

EPIC: mastering the MOS CCD production of novel open phase electrodes and device packaging, the novel PN technology CCD design and production,  developing robust algorithms for  data handling and event reduction and recognition

RGS: mastering the back-Illuminated CCD production, developing a grating box production to ensure location and stability, developing robust algorithms for  data handling and event reduction and recognition

OM : One design involving independent Red and Blue channels (the former subsequently discarded), high voltage power supply, Optical coupling of fast CCD detector readout and subsequent event centroiding algorithms

For all instruments the timely manufacturing of flight representative CCD devices with 4 different technology variants to match the different instrument needs (and readout electronics) to allow radiation damage effects testing in order to feed back into the design and production of flight devices and/or electronics, was problematic.

As with the X-ray telescope design and manufacturing, on the instrument side these significant challenges were compounded by the integration of the miscellaneous instrument models on the spacecraft and there were frequent changes in the timeline. Even when the spacecraft was already in French Guyana, being prepared for launch, instrument level swaps were necessary.

While it is not intended to describe how individual problems were eventually solved, it is very clear that the largely unknown radiation background, radiation dose and mission design lifetime played a very important role in the engineering requirements. Also a number of items were not at the required TRL (Technology Readiness Level) at the beginning of the C/D phase. Most of these problems were overcome with a staggered model delivery approach, which could be accommodated by a flexible approach in industry.

Finally, shortly before launch, the project were informed by the Chandra ACIS team (which had launched some 5 months prior to XMM-Newton) of a significant radiation problem of unknown nature, that they had encountered in orbit. In order to not delay (but also not risk) the mission a quick analysis, with extensive support from the Chandra teams, was carried out. This analysis identified the source of the problem as low energy protons which could damage the exposed surface of the CCDs. In the case of XMM-Newton, the presence of the nested X-ray telescope mirrors and the grating arrays, behind the telescopes housing the EPIC-MOS and RGS cameras, was proven by analysis (using GEANT aka *Geometry and Tracking* and some dedicated measurements) to be a very effective 'filter' in reducing any soft proton fluence. As the EPIC detectors behind the telescope not carrying an RGS grating box were much less prone to radiation damage (being back-illuminated), the mission was quickly declared to not suffer from the problem experienced by Chandra. This was verified to be the case in orbit, and radiation damage effects have been limited to the accumulation of non-ionising does from solar flare protons.

## 5 - In-orbit performance

The in-orbit performance is described in detail in the XMM-Newton Users' Handbook [18] and the three calibration status documents and,. except for the areas indicated below, matches quite well with the pre-launch predictions based on extensive simulations. Those simulations were based on a calibration and metrology based ray tracing code (SciSIM) developed especially for validating the expected scientific performance, generating the initial users' handbook and for testing the early Beta versions of the XMM-Newton Science Analysis Software (SAS).

Marked deviations from the pre-launch calibrations and modelling were observed for the RGS high energy (> 1.8 keV) effective area and EPIC high energy (> 5 keV) detector background. Where the origin of the former is not clearly understood, and the latter is most likely caused by the limited pre-launch knowledge of the cosmic ray and other background elements in an elliptic orbit like that of XMM-Newton. The EPIC PN camera response proved difficult to calibrate due to changes in some fabrication parameters been the nominally calibrated FM detector, and the eventually flown Flight Spare. In addition the OM has suffered from ghosting features due to unwanted reflections on insufficiently baffled support rings from off-axis bright sources. This was attributed omissions in optical designs when the system design was rapidly overhauled to accommodate the deletion of the red channel.

Only a very limited number of components on-board XMM-Newton failed to date. The most significant ones have led to the loss of one EPIC-MOS CCD (most likely due to a micro-meteorite) and two RGS CCD chains. Due to the use of redundancy the impact of these is scientifically very limited.

## 6 - Overall impact of XMM-Newton and size of its scientific community

The majority of XMM-Newton's observing time is made available to the astronomical community by the traditional route of Announcements of Opportunity, followed by peer review. The ten XMM-Newton "Calls for Observing Proposals" have resulted in an over-subscription of the available observing time by a factor of at least 6.5. Each call typically involves 1400 individual astronomers, which is approximately 15% of all professional astronomers worldwide. At the writing of this report more than 2780 articles based on XMM-Newton data have been published in refereed journals, at a rate which currently exceeds 300 per year. XMM-Newton publications are highly cited, about 40% of its publications belong to the top category of the 10% most cited astronomical articles. The importance of XMM-Newton is now widely recognized in the scientific world at large, beyond the X-ray and astronomical community. For illustration, Nature completed its cycle of review articles honouring the International Year of Astronomy, with a description of XMM-Newton's and Chandra's achievements [19].

## 7 - Science highlights

XMM-Newton continuously breaks new ground in many areas of science often unanticapted when the mission parameters were conceived. The following highlights illustrate the breadth of science topics to which XMM-Newton has brought significant insights:

Observations with XMM-Newton of ultimately proved charge exchange reactions between highly charged heavy ions in the solar wind - and cometary gas as cause of X-ray emission [20] XMM-Newton observations of Jupiter and Saturn have allowed a detailed understanding of the X-ray emission of the planet aurorae and magnetospheres [21,22] XMM-Newton analysis of pre-main sequence stars in molecular clouds have been instrumental in understanding the roles of feedback of the X-ray emission on the circumstellar material as well as for the fate of the circumstellar disks and the formation of planets out of these disks [ 23] Another highlight of stellar research came with the first detection of a cyclic X-ray variation in a solar type star (HD 81809); it is synchronised with the star-spot cycle, similar to that of the Sun [24 ]

The discovery of diffuse X-ray emission from a million degree gas pervading the Orion nebula came as a surprise [25], showing that a handful of massive stars is enough to generate an extensive hot bubble through the accumulation of stellar wind losses, and that bubbles of hot gas may be far more common than previously realised.

The detection of faint, hard X-ray emission from a Wolf-Rayet star WR142, sheds new light on the properties of Wolf-Rayet stellar winds and on the last stages of massive star evolution [26].

A discovery with important cosmological implications was the recognition of a new class of type Ia SN based on the emission of the Fe L-shell lines [27]. Thanks to XMM-Newton spectra, accurate relative elemental abundances could be measured in several supernovae remnants (SNR), which in turn enabled detailed tests of type Ia SNe explosion models. The prototype of shell supernova remnants, SN 1006, was compared with radio observations and led to the identification of the non-thermal synchrotron emission and showed that the magnetic field is amplified where electron acceleration is efficient [28].

Measurements of the cooling curves of transient, thermally emitting, neutron stars has allowed strong constraints on the equation of state of dense nuclear matter to be set [29].
.

XMM-Newton serendipitous discovery of the X-ray source HLX-1 in the spiral galaxy ESO 243-49 yielded the strongest case thus far for the existence of intermediate-mass black hole setting a lower limit on its mass of 500 $M_o$ [30]. Observation of the active galaxy RE J1034+396 yielded the first detection of quasi-periodic oscillations in a super-massive black hole (SMBH) [31] whose period matches that for completing one revolution in the last stable orbit around a 10 million solar mass black hole. XMM-Newton produced the first detection of a relativistically distorted iron L emission line in an AGN [32] Intensity variations of the iron L line were found to lag those of the direct X-ray continuum by 30 s, providing a direct measure of the inner radius of the accretion disk which can in turn be used to infer the mass of the black-hole, 7 million solar masses.. Further AGN observations have yielded direct measurements of the total power released by AGN winds, and provided strong evidence that these play an important role in controlling the formation of structure in the Universe [33]

The AGN feedback scenario physically connects the three main extragalactic object classes, i.e. galaxies, clusters of galaxies and AGNs, and explains their growth and evolution over cosmological times. The key measurement that the absence of a strong cooling flow for low temperatures in three clusters of galaxies was an early result of XMM-Newton. [34] Feedback is also associated with the ICM enrichment via supernovae driven galactic winds. Accurate abundances measurements provided strong new constrains on the relative abundances of different supernovae explosion [36]

A large sample of groups of galaxies in the COSMOS 2 $deg^2$ field was used to extend the validity range of the $M_H$- Lx relation to smaller groups and higher redshifts, providing an important tool for cosmology and the study of structure evolution in the Universe. [35]

Discovery of CL J1414+0856 is the most distant mature cluster of galaxies with z=2.07 [37]. and with a mass comparable with Virgo demonstrates that such objects can form significantly early in the Universe . XMM-Newton also uncovered the first evidence for the long sought "cosmic web", the dilute and warm-hot intergalactic medium , in a filament detected between the pair of clusters Abell 222 & 223 [38].

The combination of XMM-Newton and HST data and ground-based redshifts has resulted in the first 3-dimensional map of Dark Matter across the COSMOS field [39] and extending back in time by about six billion years. [40]

The 2XMM catalogue [41] currently contains 353,000 X-ray detections of 263,000 individual sources, making it the largest X-ray catalogue ever, while the X-ray Slew Catalogue [42] contains 11,400 sources detected while XMM-Newton was manoeuvring between targets. It covers 48% of the entire sky.

The observatory designed for 5 years lifetime continues to operate well into its second decade with minimal degradation in performance The prospects for continued productive life augurs well for the rich harvest of XMM-Newton science to be continued and to be enhanced with co-operative progranmmes with other new facilities (e.g. Herschel Planck, LOFAR, ALMA and Fermi ).

## 8 - Acknowledgements

It is with great please we acknowledge all the teams and individuals who have contributed to the success of the XMM-Newton project. Particular thanks go to the Principle Investigator teams who provided the instruments, telescope calibrations and data analysis systems, and the operations staff in the Mission and Science Operations Centres.

## 9 - References

Biographies: (photos not available)

David Lumb obtained a PhD from Leicester University 1983 with the topic of "CCDs for X-ray Astronomy", and then continued to develop CCD technology for XMM-Newton and other missions. He was program manager for space missions at Penn State University (1990-1994). He joined ESA as XMM Instrument and Calibration Team Leader (1994-2001). He moved to ESA's Advanced Technologies and Concepts Office to work on diverse instruments and optics for future astronomy missions, (2002-2009). Currently he is the ESA Study Scientist for IXO, Athena, LOFT and Astro-H missions (2009-2011)

Fred Jansen obtained a PhD in astrophysics at Leiden University 1988 where he worked on development of CCD's for RGS instrument on XMM-Newton. Joined ESA in 1995 and was XMM-Newton project scientist and mission manager (1997-2004 and 2000-2006 respectively) and Gaia project scientist for 1 year (2006). Currently Mars Express and Venus Express mission manager and science ground segment development manager for BepiColombo.

Norbert Schartel completed his Diploma in Physics in June 1990 at the Technische Hochschule, Darmstadt, specialising in General Relativity. He then received his PhD in Physics from the Ludwig-Maximilians University in Munich for the research on ROSAT X-ray spectra of quasars he performed at the Max-Planck-Institut für extraterrestrische Physik in Garching. He joined ESA in 1994 as a resident astronomer at ESAC, where he worked on the International Ultraviolet Explorer, a space observatory for ultraviolet spectrometry which was launched in 1978. Norbert Schartel joined the XMM-Newton team in 1998 being appointed Project Scientist in 2004.